\documentclass{aa}
\usepackage[varg]{txfonts}

\usepackage{graphicx}
\usepackage{natbib}
\bibpunct{(}{)}{;}{a}{}{,} % to follow the A&A style

\begin{document}

%\setlength{\droptitle}{-4\baselineskip} % Move the title up
%
%\pretitle{\begin{center}\Huge\bfseries} % Article title formatting
%\posttitle{\end{center}} % Article title closing formatting
%\title{A fast migration rate for Titan can explain Iapetus' orbit} % Article title
%\author{%
%%\textsc{William POLYCARPE, Valery Lainey, Alain Vienne, Benoit Noyelles, Melaine Saillenfest, Nicolas Rambaux}
%William POLYCARPE, Valery Lainey, Alain Vienne,\\
%\\ Benoit Noyelles, Melaine Saillenfest, Nicolas Rambaux
%%\thanks{A thank you or further information} \\[1ex] % Your name
%%\normalsize Observatoire de Paris // IMCCE \\ % Your institution
%%\normalsize \href{mailto:william.polycarpe@obspm.fr}{william.polycarpe@obspm.fr} % Your email address
%%\and % Uncomment if 2 authors are required, duplicate these 4 lines if more
%%\textsc{Jane Smith}\thanks{Corresponding author} \\[1ex] % Second author's name
%%\normalsize University of Utah \\ % Second author's institution
%%\normalsize \href{mailto:jane@smith.com}{jane@smith.com} % Second author's email address
%}
%\date{} % Leave empty to omit a date
%\renewcommand{\maketitlehookd}{%
%\title{A fast migration rate for Titan can explain Iapetus' orbit}
\title{Strong tidal energy dissipation in Saturn at Titan's frequency as an explanation for Iapetus orbit}
\author{William Polycarpe\inst{1}
\and Melaine Saillenfest\inst{1}
\and Valéry Lainey\inst{2}
\and Alain Vienne\inst{1}
\and Benoît Noyelles\inst{3}
\and Nicolas Rambaux\inst{1}}

\institute{IMCCE, Observatoire de Paris, PSL Research University, CNRS, UPMC Univ. Paris 06, Univ. Lille.
\and Jet Propulsion Laboratory, California Institute of Technology
\and Namur Institute for Complex Systems (naXys), University of Namur}

\date{Received date/Accepted date}
\abstract{Natural satellite systems present a large variety of orbital configurations in the solar system. While
some are clearly the result of known processes, others still have largely unexplained eccentricity and
inclination values. Iapetus, the furthest of Saturn’s main satellites, has a still unexplained 3\% orbital
eccentricity and its orbital plane is tilted with respect to its local Laplace plane (8 degrees of free inclination). On the other hand, astrometric measurements of saturnian moons have revealed high tidal migration rates , corresponding to a quality factor $Q$ of Saturn of around 1600 for the mid-sized icy moons.}{We show how a past crossing of the 5:1 mean motion resonance between Titan and Iapetus may be a plausible scenario to explain Iapetus' orbit.}{We have carried out numerical simulations of the resonance crossing using an N-Body code as well as using averaged equations of motion. A large span of migration rates were explored for Titan and Iapetus was started on its local Laplace plane (15 degrees with respect to the equatorial plane) with a circular orbit.}{The resonance crossing can trigger a chaotic evolution of the eccentricity and the inclination of Iapetus. The outcome of the resonance is highly dependent on the migration rate (or equivalently on $Q$). For a quality factor $Q$ of over around 2000, the chaotic evolution of Iapetus in the resonance leads in most cases to its ejection, while simulations with a quality factor between 100 and 2000 show a departure from the resonance with post-resonant eccentricities spanning from 0 up to 15\%, and free inclinations capable of reaching 11 degrees. Usually high inclinations come with high eccentricities but some simulations (less than 1\%) show elements compatible with Iapetus' current orbit}{In the context of high tidal energy dissipation in Saturn, a quality factor between 100 and 2000 at the frequency of Titan would bring Titan and Iapetus into a 5:1 resonance, which would perturb Iapetus' eccentricity and inclination to values observed today. Such rapid tidal migration would have avoided Iapetus' ejection around 40 to 800 million years ago.}


\titlerunning{A fast migrating Titan can explain Iapetus' orbit}
\authorrunning{William Polycarpe et al.}

\maketitle

\section{Introduction}
Mean motion resonances in planetary systems have been the subject of many studies throughout the last century, up to the present day. It has been established that mean motion ratios that are close to rational values are very unlikely to be the effect of randomness \citep{1954MNRAS.114..232R}, being instead the result of a series of mechanisms involving satellite migration and subsequent capture into stable mean motion resonant configurations \citep{1965MNRAS.130..159G}. The system formed by Saturn and its eight major satellites (Mimas, Enceladus, Tethys, Dione, Rhea, Titan, Hyperion, and Iapetus) is a unique system in which one can find several of its moons evolving in a resonance.
We have Mimas and Tethys, locked in a 4:2 mean motion resonance, responsible for their growth in inclination; Enceladus and Dione whose periods are commensurable, yielding a 2:1 resonance that explains the eccentricity of Enceladus (and is a direct cause of the heating of Enceladus); and finally Titan and the small moon Hyperion, which keep their orbital periods in a 4:3 resonance, Titan being known to drive Hyperion's orbit thanks to this resonance \citep{1974AJ.....79...61C}.

Today it is believed that tides from satellites acting on the planet are responsible for satellite resonances, the latter preserving the commensurability as the satellites continue evolving outwards. The general theory would relate the growth of the semi-major axis of the satellite to the dissipation acting inside the planet. Such a theory is applicable to solid bodies like our Earth but it is not straightforward to apply it to gaseous bodies such as giant planets or stars. However, scientists usually keep the dissipative approach and relate  satellite migration to the quality factor $Q$, which measures the energy dissipation inside the planet.

However, the Saturn system has recently been targeted by astrometry. First, the spacecraft Cassini, in orbit around Saturn between 2004 and 2017, was a key instrument for the measurements of the positions of Saturn's satellites throughout its mission \citep{2018A&A...610A...2C}. Many Earth-based observations of this system were also made in the last hundred years. Making use of all these measurements, \cite{2017Icar..281..286L} fitted an average quality factor $Q$ of around 1600 at the orbital frequency of the icy moons Enceladus, Tethys, and Dione, which is more than an order a magnitude smaller than the theoretical  value attributed to Saturn's dissipation by \cite{1966Icar....5..375G}. In addition, the quality factor at Rhea's frequency was found to be around 300, not only confirming a fast tidal recession for this moon but also suggesting that the quality factor could be a frequency-dependent parameter. These high dissipation values may be explained by viscous friction in a solid core in Saturn \citep{2012A&A...541A.165R}, resulting in a fast expansion of the orbits.

Assuming high $Q$ at Mimas's frequency \citep{1966Icar....5..375G}, some authors suggested that $Q$ could still be around 2000 at Titan's frequency (\cite{1973AJ.....78..338G}; \cite{1974AJ.....79...61C}). A more recent theory \citep{2016MNRAS.458.3867F} gave a explanation for the low and frequency-dependent values of $Q$ found in \cite{2017Icar..281..286L}. Applied to Titan, $Q$ would be around 25 \citep{2016MNRAS.458.3867F}.

These fast tidal recessions allow for the possibility of past resonance crossings. If Titan had been migrating very rapidly in its history ($Q$ < $\approx~14\,000$), it would have crossed a 5:1 mean motion resonance with Iapetus. Therefore, in this study, we explore several values of $Q$ at Titan's frequency with a particular emphasis on low values (2-5000). We simulate a 5:1 mean motion resonance crossing with Iapetus over a few million to hundreds of millions of years. Iapetus is today on an eccentric orbit ($e_{I}\approx 0.03$) and its orbital plane has an 7.9 degree tilt with respect to its local Laplace plane (Figure \ref{fig:tenbody}). If created from its circumplanetary disk, the tilt should have been null \citep{2014AJ....148...52N}. Some authors have proposed some plausible scenarios to explain these values. \cite{1981Icar...46...97W} invoked a fast dispersal of the circumplanetary disk from which satellites were formed. Iapetus would have evolved to the orbit we see today if the disk was dispersed in around 10\textsuperscript{2} or 10\textsuperscript{3} years, which is a fast timescale for this process to happen. Also, in the context of the early solar system instability \citep{2005Natur.435..459T}, \cite{2014AJ....148...52N} argue that Iapetus could have been excited by close encounters between an ice giant and the system of Saturn. However, we looked for an alternative scenario where the 5:1 resonance would be responsible for its orbital behaviour.

We first introduce the general theory of a satellite orbiting on or close to its Laplace plane. Then, after explaining the different methods used for this research, we describe the results of several simulations made numerically with an N-body code, or using averaged equations of motion. Finally, the article ends with some discussion on the method and the conclusions.
\section{The Laplace plane}
Iapetus' orbital plane is actually tilted with respect to several important planes. First, due to the relatively large distance of the satellite from the planet, the effect coming from the Sun is more pronounced than for the other satellites. Iapetus does not orbit in the planet's equator. Its orbital plane secularly oscillates around a specific plane called the Laplace plane. The latter is a fixed plane with an inclination of about 15 degrees with respect to the equator. Orbiting close to it, a satellite would also have its ascending node oscillating around a fixed value, corresponding to the Sun's ascending node. It is defined as an equilibrium between the pull of the Sun and the flattening effect of the planet and the inner satellites.
\begin{figure}
\resizebox{\hsize}{!}{\includegraphics{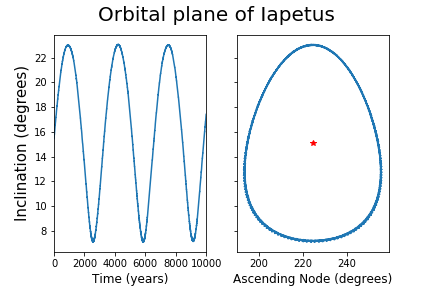}}
\caption{Motion of Iapetus plane over $10\,000$ years. Both the inclination and the longitude of the ascending node oscillate around the fixed values of the Laplace plane with a period of around 3200 years. Here, and for the rest of this work, the reference plane is the equator of Saturn. The figure comes from a simulation done with all the major satellites of Saturn (Mimas, Enceladus, Tethys, Dione, Rhea, Titan, Hyperion, Iapetus), the Sun, and the effect of the flattening of Saturn. The angular elements of the Laplace plane are approximately 15 degrees for the inclination and 225 degrees for the ascending node. The initial conditions are those of the J2000 epoch taken from JPL ephemeris, Horizon (SAT 389.14).}
\label{fig:tenbody}
\end{figure}
The orbit of Iapetus is tilted with respect to this plane with an angle of about 7.9 degrees. On top of this, its eccentricity is around $0.03$. We consider that both values still need an explanation. We assume that a satellite would be created on its Laplace plane after being formed from a circumplanetary disk (\cite{2014AJ....148...52N}).

\begin{figure}
\resizebox{\hsize}{!}{\includegraphics{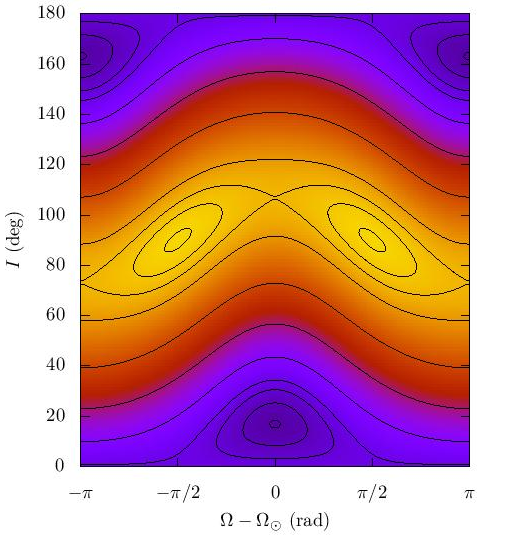}}
\caption{Level curve of the Hamiltonian from Eq. \ref{rlp}. Orbital plane elements will follow the solid lines. One can distinguish four different kinds of equilibrium: one for retrograde orbits, two for polar orbits, and the one we are interested in, for relatively low inclination. The latter gives us $\Omega=\Omega_{\odot}$ and both ascending nodes of Iapetus and Titan oscillate around it. Equation \ref{tan2i} and Table \ref{table:j2pt} give an inclination of the local Laplace plane of Titan around 0.62 degrees}
\label{table:Hlevel}
\end{figure}

To set the general equations for the Laplace plane, we start with the perturbing acceleration of a body $j$ acting upon a body $i,$
\begin{equation}
\ddot{{\bf r}}_{i}+G\left(m_{S}+m_{i}\right)\frac{{\bf r}_{i}}{r_{i}^{3}}=Gm_{j}\left(\frac{{\bf r}_{j}-{\bf r}_{i}}{r_{ij}^{3}}-\frac{{\bf r}_{j}}{r_{j}^{3}}\right)
\label{mutual}
,\end{equation}
where ${\bf r}_{i}$ denotes the position vector of the $i^{th}$ satellite, with mass $m_{i}$,  and the unbold $r_{i}$ its magnitude. So $r_{ij}$ stands for the separation between satellites $i$ and $j$, $|{\bf r}_{i}-{\bf r}_{j}|$. The mass of Saturn is denoted by $m_{S}$ and $G$ stands for the gravitational constant. The right-hand side of this equation can be rewritten $\nabla_{i}R_{j}$, where $R_{j}$ stands for the disturbing function of the body $j$ acting on $i,$
\begin{equation}
R_{j}=Gm_{j}\left(\frac{1}{r_{ij}}-\frac{{\bf r_{i}}.{\bf r_{j}}}{r_{j}^{3}}\right).
\end{equation}
In our case, the reference frame is centred on Saturn and the Sun acts as an external satellite around the planet. The disturbing function for the Sun can therefore be expanded in terms of Legendre polynomials \citep{murray2000solar} \citep{brouwer1961methods},
\begin{equation}
R_{\odot}=\frac{Gm_{\odot}}{r_{\odot}}\sum_{l=2}^{\infty}\left(\frac{r_{i}}{r_{\odot}}\right)^{l}P_{l}\left(\cos(\psi)\right).
\label{rsunc}
\end{equation}
Here the subscript $i$ stands for any satellite and $\psi$  denotes the angle between the position vectors of the Sun and the satellite, verifying the relation ${\bf r_{i}}.{\bf r_{\odot}}=r_{i}r_{\odot}\cos(\psi)$. After some quick manipulations\footnote{$Gm_{\odot}$ should be replaced rigorously by $\frac{n_{\odot}^{2}a_{\odot}^{3}}{1+\frac{m_{S}}{m_{\odot}}}$, but $\frac{m_{S}}{m_{\odot}} \approx 3\times 10^{-4}$ so $Gm_{\odot}\approx n_{\odot}^{2}a_{\odot}^{3}$.} \citep{brouwer1961methods}, terms in series \ref{rsunc} appear proportional to $\alpha^{l}$, $\alpha$ being the semi-major axis ratio, which is small in the case of planetary satellites.\footnote{The Titan-Sun semi-major axis ratio is $8.5\times 10^{-4}$ and the one concerning Iapetus and the Sun is $2.5\times 10^{-3}$.} Therefore, we limit ourselves to the first term in the series, $l=2,$
\begin{equation}
R_{\odot}=n_{\odot}^{2}a_{i}^{2}\left[\left(\frac{r_{i}}{a_{i}}\right)^{2}\left(\frac{a_{\odot}}{r_{\odot}}\right)^{3}\left(-\frac{1}{2}+\frac{3}{2}\cos^{2}(\psi)\right)+O(\alpha^{2})\right]
.\end{equation}
Using several expansions developed in \cite{murray2000solar}, and after averaging over the mean longitudes of the satellite and the Sun, we obtain eight different arguments
\begin{table*}
\caption{Terms coming from the disturbing function of the Sun acting on both satellites.}
\label{table:solarterms}
\centering
\begin{tabular}{c|c}
\hline
\hline
Cosine argument & Term\\
\hline
$\emptyset$&$\frac{1}{32}\left[8+12e_{\odot}^{2}+15e_{\odot}^{4}\right.\left.+6e_{i}^{2}\left(2+3e_{\odot}^{2}\right)\right]\left(1-6s_{i}^{2}+6s_{i}^{4}\right)\left(1-6s_{\odot}^{2}+6s_{\odot}^{4}\right)$\\
\hline
$2\Omega_{i} - 2\Omega_{\odot}$&$\frac{3}{8}\left[8+12e_{\odot}^{2}+15e_{\odot}^{4}\right.\left.+6e_{i}^{2}\left(2+3e_{\odot}^{2}\right)\right]s_{i}^{2}\left(1-s_{i}^{2}\right)s_{\odot}^{2}\left(1-s_{\odot}^{2}\right)$\\
\hline
$\Omega_{i} - \Omega_{\odot}$&$\frac{3}{8}\left[8+12e_{\odot}^{2}+15e_{\odot}^{4}\right.\left.+6e_{i}^{2}\left(2+3e_{\odot}^{2}\right)\right]s_{i}\sqrt{1-s_{i}^{2}}\left(1-2s_{i}^{2}\right)s_{\odot}\sqrt{1-s_{\odot}^{2}}\left(1-2s_{\odot}^{2}\right)$\\
\hline
$2\Bar{\omega}_{i} - 2\Omega_{\odot}$&$\frac{15}{8}e_{i}^{2}\left(2+3e_{\odot}^{2}\right)\left(1-s_{i}^{2}\right)^{2}s_{\odot}^{2}\left(1-s_{\odot}^{2}\right)$\\
\hline
$4\Omega_{i} - 2\Omega_{\odot} - 2\Bar{\omega}_{i}$&$\frac{15}{8}e_{i}^{2}\left(2+3e_{\odot}^{2}\right)s_{i}^{4}s_{\odot}^{2}\left(1-s_{\odot}^{2}\right)$\\
\hline
$3\Omega_{i} - \Omega_{\odot} - 2\Bar{\omega}_{i}$&$\frac{15}{4}e_{i}^{2}\left(2+3e_{\odot}^{2}\right)s_{i}^{3}\sqrt{1-s_{i}^{2}}s_{\odot}\sqrt{1-s_{\odot}^{2}}\left(1-2s_{\odot}^{2}\right)$\\
\hline
$\Omega_{i} + \Omega_{\odot} - 2\Bar{\omega}_{i}$&$-\frac{15}{4}e_{i}^{2}\left(2+3e_{\odot}^{2}\right)s_{i}\left(1-s_{i}^{2}\right)^{3/2}s_{\odot}\sqrt{1-s_{\odot}^{2}}\left(1-2s_{\odot}^{2}\right)$\\
\hline
$2\Bar{\omega}_{i} - 2\Omega_{i}$&$\frac{15}{8}e_{i}^{2}\left(2+3e_{\odot}^{2}\right)s_{i}^{2}\left(1-s_{i}^{2}\right)\left(1-6s_{\odot}^{2}+6s_{\odot}^{4}\right)$\\
\hline
\end{tabular}
\end{table*}
with $s_{i}=\sin(\frac{i_{i}}{2}).$
Then, the satellites  also undergo the effect of the flattening of Saturn; the disturbing function related to this effect can be written
as\begin{equation}
R_{J_{2}}=\frac{G\left(m_{S}+m_{i}\right)Rp^{2}J'_{2}\left[1+3\cos(2i)\right]}{8a^{3}\left(1-e^{2}\right)^{3/2}}
%\ddot{{\bf r}}_{i}=\frac{3}{2}R_{S}^{2}J'_{2}G\left(m_{S}+m_{i}\right)\frac{1}{r^{7}_{i}}\times \left( \begin{array}{c}x_{i}(5z_{i}^{2}-r_{i}^{2}) \\ y_{i}(5z_{i}^{2}-r_{i}^{2}) \\ z_{i}(5z_{i}^{2}-3r_{i}^{2})\end{array}\right)
\label{rj2}
,\end{equation}
with $R_{p}$ the equatorial radius of Saturn. Here, it is important to underline that $J_2'$ not only corresponds to the zonal harmonic of Saturn, but to a general flattening seen by a satellite, made of the proper flattening of the planet and by the secular effect of the interior satellites. We define \citep{2009AJ....137.3706T}
\begin{equation}
J'_{2}=J_{2}+\frac{1}{2}\sum\left(\frac{a_{i}}{R_{S}}\right)^{2}\frac{m_{i}}{m_{S}}
\label{j2p}
.\end{equation}
As a consequence, Titan and Iapetus do not see the same "flattening". Titan  undergoes the flattening of the planet and the inner icy satellites (Mimas to Rhea), which do not participate much, while it  plays a major role in the $J_{2}'$ for Iapetus.
\begin{table}
\caption{Flattening parameters seen by both satellites. The contribution coming from the inner icy satellites is small and equals roughly $2.1\times10^{-4}$. The $J_{2}$ harmonic coefficient of Saturn is chosen to be around $0.01629$ (from SAT 389.14).}
\label{table:j2pt}
\centering
\begin{tabular}{c|c}
\hline\hline
Satellite & $J_{2}'$\\
\hline
Titan&0.01650\\
\hline
Iapetus&0.0650\\
\hline
\end{tabular}
\end{table}
We form the disturbing function formed by those two effects,
\begin{equation}
R_{LP}=R_{\odot}+R_{J_{2}}
\label{rlp}
,\end{equation}
and plug it into the Lagrange Planetary Equations
\begin{equation}
\frac{di}{dt}=-\frac{1}{na^{2}\sqrt{1-e^{2}}}\left(\tan\left(\frac{i}{2}\right)\frac{\partial R_{LP}}{\partial \Bar{\omega}}+\frac{1}{\sin(i)}\frac{\partial R_{LP}}{\partial \Omega}\right)
,\end{equation}
\begin{equation}
\frac{d\Omega}{dt}=\frac{1}{na^{2}\sqrt{1-e^{2}}\sin(i)}\frac{\partial R_{LP}}{\partial i}
.\end{equation}
A satellite orbiting on its local Laplace plane will see its orbital plane fixed in time, meaning that the rates of the ascending node and the inclination are null. The local Laplace plane orbital elements therefore verify
\begin{equation}
\left\{\begin{array}{ccl}
\frac{di_{LP}}{dt}&=&0\\
\\
\frac{d\Omega_{LP}}{dt}&=&0
\end{array}\right.
\label{lpequi}
,\end{equation}
which possess several solutions (see Figure \ref{table:Hlevel}).
The solution we are interested in imposes
\begin{equation}
\left\{\begin{array}{ccl}
\epsilon_{J_{2}}\sin(2i_{LP})+\epsilon_{\odot}\sin(2i_{LP}-2i_{\odot})&=&0\\
\Omega_{LP}-\Omega_{\odot}&=&0
\end{array}\right.
\label{lpiap}
.\end{equation}
The inclination and ascending node of the Laplace plane are those verifying formula \ref{lpiap}. We see in Fig. \ref{fig:tenbody} that the ascending node oscillates around a fixed value. This value is actually that of the Sun, $\Omega_{\odot}$, which was around 225 degrees for this specific simulation. In the rest of this study we have set $\Omega_{\odot}=0$. The equilibrium inclination, $i_{LP}$,  can be found by rewriting the first equation of formula \ref{lpiap},
\begin{equation}
\tan\left(2i_{LP}\right)=\frac{\sin(2i_{\odot})}{\frac{\epsilon_{J_{2}}}{\epsilon_{\odot}}+\cos\left(2i_{\odot}\right)}
\label{tan2i}
,\end{equation}
\begin{figure}
\resizebox{\hsize}{!}{\includegraphics{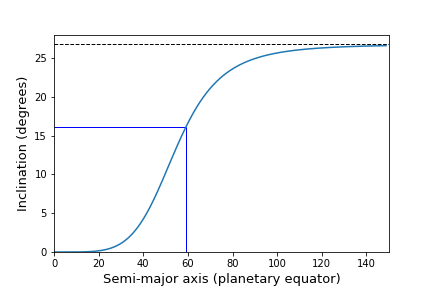}}
\caption{Tilt of Iapetus' local Laplace plane to Saturn's equator as a function of Iapetus' semi-major axis. Here $J_{2}'$ includes the secular effect of Titan. Equation \ref{tan2i} was used to produce this graph and it gives a Laplace plane inclination of around 16 degrees for the semi-major axis of Iapetus today (0.02381 AU). These numbers are represented by the blue lines. This plot also shows that a satellite would usually orbit close to the planet's equator if its orbit is close to it and would see its orbital plane getting close to the ecliptic as its semi-major axis gets higher. Iapetus lies in-between those two cases.}
\end{figure}
where
\begin{equation}
\epsilon_{J_{2}}=Gm_{S}\left(1+\frac{m_{i}}{m_{S}}\right)\frac{16J_{2}R_{p}^{2}}{\left(1-e_{i}^{2}\right)^{3/2}}
\end{equation}
and
\begin{equation}
\begin{split}
\epsilon_{\odot}&=Gm_{\odot}\left(1+\frac{m_{S}}{m_{\odot}}\right)8a_{i}^{2}\alpha_{i, \odot}^{3}\\
&\times \left(1+\frac{3}{2}e_{\odot}^{2}+\frac{15}{8}e_{\odot}^{4}+\frac{1}{4}e_{i}^{2}\left(2+3e_{\odot}^{2}\right)\left(3-5\cos(2\Bar{\omega}_{i})\right)\right)
\end{split}
,\end{equation}
so that for $\frac{m_{S}}{m_{\odot}}<<1$ and $\frac{m_{i}}{m_{S}}<<1$
\begin{equation}
\frac{\epsilon_{J_{2}}}{\epsilon_{\odot}}\approx2\frac{m_{S}}{m_{\odot}}\frac{J_{2}R_{p}^{2}a_{\odot}^{3}}{a_{i}^{5}}F(e_{\odot}, e_{i})
.\end{equation}
The function $F(e_{\odot}, e_{i})$ contains the dependency on both eccentricities,
\begin{equation}
\begin{split}
F(e_{\odot}, e_{i})&=\frac{1}{\left(1-e_{i}^{2}\right)^{3/2}}\\
&\times \frac{1}{\left(1+\frac{3}{2}e_{\odot}^{2}+\frac{15}{8}e_{\odot}^{4}+\frac{1}{4}e_{i}^{2}\left(2+3e_{\odot}^{2}\right)\left(3-5\cos(2\Bar{\omega}_{i})\right)\right)}
\end{split}
\label{fee}
.\end{equation}
%The dependency on $\omega_{i}$ can be neglected and dropped for convenience.
Eccentricities do not play a major role in the equilibrium if they stay small. However, if we consider the action of other planets, the eccentricity of Saturn (or the Sun in our model) will change, changing slightly the position of the Laplace plane.

We have derived here the equation for the inclination of the Laplace plane. Equation \ref{tan2i} can also be found in \cite{2009AJ....137.3706T} and in \cite{2014AJ....148...52N}. For Titan and Iapetus, it is a forced plane around which their orbit will evolve. In our case, Iapetus is assumed to have formed on its Laplace plane, therefore we chose it as the initial orbital plane of the satellite in our simulations.

\section{Tides}
The start of this research was motivated by the low values of the quality factor $Q$ found at the frequency of the inner icy satellites in \cite{2012ApJ...752...14L} and \cite{2017Icar..281..286L}. Furthermore, \cite{2016MNRAS.458.3867F} published a theory in which those dissipative values  matched the ones found by astrometry. Fuller's model also predicts a quality factor of around 25 for Titan if we consider a tidal timescale of around two billion years (second equation in Fuller's paper), as was chosen for other satellites in the paper. Fuller's model predicts that resonance locking can be reached due to the internal structural evolution of the planet. Moons are then "surfing" on dissipation peaks and migrate outwards at rates comparable to the ones found by astrometry. We  assume relatively high migration rates for Titan by choosing a quality factor spanning from 2 to around 5000, but also higher values were tested to be able to make a full comparison. The following tidal acceleration was implemented in our software \citep{2007JGRE..11212003E}:
\begin{equation}
\ddot{{\bf r}}_{i}=-\frac{3k_{2}R_{p}^{5}G\left(m_{S}+m_{i}\right)m_{i}}{r_{i}^{10}m_{S}}\times\Delta t\left(2{\bf r}_{i}\left({\bf r}_{i}.{\bf v}_{i}\right)+r_{i}^{2}\left({\bf r}_{i}\times {\bf \Omega}_{S}+{\bf v}_{i}\right)\right)
\label{tides}
,\end{equation}
with
\begin{equation}
\Delta t=T\arctan(1/Q)/2\pi
\label{deltat}
\end{equation}
and
\begin{equation}
T=\frac{2\pi}{2\left(\Omega-n\right)}
.\end{equation}
Here $k_{2}$ denotes the second order Love number, $R_{p}$ is the equatorial radius of Saturn, $G$ the gravitational constant, $m_{S}$, Saturn's mass, $m_{i}$ the satellite's mass, ${\bf r}_{i}$ its positional vector and its norm $r_{i}$, ${\bf v}_{i}$ its velocity vector, and ${\bf \Omega}_{S}$ the rotational spin axis of the planet. For the latter, Saturn's rotational period is taken to be 0.44401 days.

This tidal equation will introduce a secular increase of Titan's semi-major axis, following
\begin{equation}
\frac{da}{dt}=\frac{3k_{2}R_{p}^{5}nm_{T}}{m_{S}a^{4}}\arctan\left(\frac{1}{Q}\right)
\label{tidessec}
,\end{equation}
and also of its eccentricity, but the change is negligible. This tidal model will not produce the same migration of satellite on a long timescale as the model proposed in \cite{2016MNRAS.458.3867F}, but here we are looking for a classical tidal model producing a constant time rate of the semi-major axis through the resonance. We underline here that tidal theories were first  produced for solid bodies. In these theories, the satellite  creates a bulge on the planet through differential acceleration, but this bulge would not align instantaneously with the satellite because of the viscoelasticity of the primary (\cite{1964RvGSP...2..661K}; \cite{1980M&P....23..185M}) . The quality factor $Q$ introduced in Eqs. \ref{deltat} and \ref{tidessec} would measure the energy dissipated inside the planet. The consequence of this lag is the triggering of angular momentum exchange between the planet spin and the orbit of the satellite. Whether this angular momentum exchange happens in the core or in the envelope of the planet is still under investigation. However, it is understood that the core can strongly contribute to the tidal acceleration of satellites due to internal dissipation and the existence of a solid inner core in Saturn has been recently reconsidered (\cite{2015A&A...573A..23R}; \cite{2014A&A...566L...9G}). The convective envelope would be subjected to matter redistribution and modification of the velocity field in the presence of a tidal potential (\cite{2012A&A...544A.132R}). This could also trigger some viscous friction and exchange of angular momentum between the planet and the satellite (\cite{1966AnAp...29..313Z}). Bodely tides formalism can also apply to fluid envelops (\cite{2012A&A...544A.133E}). Instead of using the notion of energy dissipation, one can use its counterpart in terms of semi-major axis growth with Eq. \ref{tidessec}. Unless one chooses a very low value of $Q$, the semi-major axis of Titan will not change drastically during the resonance crossing. Therefore we can plug constant values into the right-hand side of Eq. \ref{tidessec}. Rewritten in units of Titan's initial semi-major axis and million years, the semi-major axis rate would be approximately
\begin{equation}
\frac{da}{dt}=0.0105\times\arctan\left(\frac{1}{Q}\right)
\label{equ:dadtq}
.\end{equation}
For instance, for $Q=100$, Titan increases $0.01\%$ of its initial semi-major axis in a million years.
\section{Methods}
\subsection{Summary of the physical model}
All the above equations are set with the origin of the reference frame centred in Saturn, from which the $x$ and $y$ axes lie in the equatorial plane of the planet and the $z$ axis points towards the north pole. The study involves Titan and Iapetus, but in order to catch the correct dynamics, we have shown that the gravitational attraction of the Sun has to be added. It was placed on an orbit with a $26.73\,^{\circ}$ inclination, which corresponds to Saturn's obliquity. The presence of Hyperion is also important since it lies in a 4:3 mean motion resonance with Titan. We have kept track of such a dynamical configuration. The contributions coming from other satellites were merged with the flattening effect of the planet by defining an upgraded $J_{2}$ effect (Eq. \ref{j2p}). The gravity of the other planets has been neglected, but nevertheless, the action of Jupiter adds a small secular oscillation to the Laplace plane equilibrium. It will be discussed at the end of the paper. Finally, Titan would tidally migrate during the simulation, with a rate depending on the quality factor of Saturn (Eq. \ref{tidessec}). Finally, we neglect Iapetus' migration.
\subsection{Direct-Nbody integration and averaged equations}
Two different methods have been used. The first and more direct is an integration of the positions and velocities of the bodies using an implicit Gauss-Radau scheme like in \cite{1974CeMec..10...35E}, \cite{1985dcto.proc..185E}, and \cite{2015MNRAS.446.1424R}. This method, although precise, is time-consuming and needs a good amount of hardware power. Simulations involving a parameter $Q$ below 1500 were done using this approach, but for higher values of the quality factor, averaged equations of motion were used. In this approach, we used non-singular coordinates to avoid any singularity problem, as Titan orbits close to the equator and Iapetus' orbit was chosen as circular at the beginning of the integration. Effects coming from the flattening of Saturn were implemented by using partial derivatives of Eq. \ref{rj2} with respect to those coordinates and then plugged into the Lagrange Planetary Equations. Tides were added by using Eq. \ref{tidessec} and the averaged effect of the Sun is well represented by the eight terms in Table \ref{table:solarterms}. In order to get a good approximation for the mutual influence of the satellite, we chose to use the method developed in \cite{murray2000solar}. Titan acts like a internal perturber for Iapetus and Iapetus like an external one for Titan, therefore the disturbing function for those two satellites can be written as
\begin{equation}
R_{Titan}=\frac{Gm_{T}}{a_{I}}\left(R_{D}+\frac{1}{\alpha^{2}}R_{I}\right)
\label{rt}
\end{equation}
\begin{equation}
R_{Iapetus}=\frac{Gm_{I}}{a_{I}}\left(R_{D}+\alpha R_{E}\right).
\label{ri}
\end{equation}
Using this notation, Iapetus' dynamics, for instance, will be given by first taking the partial derivatives of $R_{Titan}$ with respect to Iapetus coordinates, and then plugged into the Lagrange Planetary Equations. For both Eqs. \ref{rt} and \ref{ri}, $R_{D}$ is the direct part of the disturbing function
\begin{equation}
R_{D}=\frac{a_{I}}{|{\bf r}_{T}-{\bf r}_{I}|}.
\end{equation}
Here, $R_{I}$ and $R_{E}$ are the indirect part of the disturbing function for the internal and external perturber. Each of those terms has an explicit development shown in \cite{murray2000solar} and we have made an intensive use of them. Terms appearing in the development have the general form
\begin{equation}
S(e', e, i', i, \alpha)\cos\left(j_{1}\lambda'+j_{2}\lambda+j_{3}\bar{\omega}'+j_{4}\bar{\omega}+j_{5}\Omega'+j_{6}\Omega\right),
\end{equation}
where the d'Alembert rule imposes that the sum of the integer coefficients $j_{i}$ vanishes and that by symmetry, $j_{5}+j_{6}$ is always an even number. In our case, we are studying dynamics that involve a close or exact 5:1 commensurability between the mean motion of the satellites. Therefore, we limit ourselves to two types of arguments: those with no mean longitudes appearing, which we call the secular terms, and those that have $5\lambda_{I}-\lambda_{T}$ in their expression. We call those resonant terms. Terms other than those mentioned are averaged out. For resonant terms, we have the relation 
\begin{equation}
j_{3}+j_{4}+j_{5}+j_{6}=-4
\end{equation}
imposed by the d'Alembert rule. The function $S$ depends on the eccentricities, inclinations, and the semi-major axis ratio, which relates to the cosine argument in the following way:
\begin{equation}
S=e_{I}^{|j_{3}|}e_{T}^{|j_{4}|}s_{I}^{|j_{5}|}s_{T}^{|j_{6}|}\times F\left(e_{I}, e_{T}, s_{I}, s_{T}, \alpha\right).
\end{equation}
The expansion in \cite{murray2000solar} assumes that eccentricities and inclinations are small, and a fourth order expansion would mean
\begin{equation}
|j_{3}|+|j_{4}|+|j_{5}|+|j_{6}|\leq 4
\end{equation}
anywhere in the expansion. However, in our case we have to account for the fact that Iapetus' orbital plane is far from being equatorial. On average the inclination will be 15 degrees, the tilt of the Laplace plane. Therefore $s_{I}$ is not small and the consequence is that higher powers of it have to be added to ensure a correct expansion. On the other hand, Titan's local Laplace plane is close to the equator and $s_{T}<0.01$ during its evolution. Therefore no power higher than 2 for $s_{T}$ will be considered. For eccentricities, we had to go up to fourth powers because it is a resonance of order 4.

This redefines the idea we have of the order of the expansion. The following constraints where used to truncate our expansion
:\begin{equation}
|j_{3}|+|j_{4}|+|j_{6}|\leq 4
\end{equation}
with
\begin{equation}
|j_{6}|\leq 2
\end{equation}
%and
%\begin{equation}
%|j_{3}|+|j_{4}|\leq 4
%\end{equation}
and $j_{5}$ is left unconstrained, but we have truncated the expansion in $s_{8}$ when the numerical evaluation of the disturbing term stopped showing a significant change. The resulting expansion for the direct part has 31 secular and 61 resonant terms. Here is an example of a term
\begin{equation}
\begin{split}
R_{res,19}&=e_{I}e_{T}s_{I}^{2}s_{T}^{2}\left(\Phi_{19,0}+s_{I}^{2}\Phi_{19,1}+s_{I}^{4}\Phi_{19,2}+s_{I}^{6}\Phi_{19,3}\right)\\
&\times \cos\left(5\lambda_{I}-\lambda_{T}-\bar{\omega}_{I}+\bar{\omega}_{T}-2\Omega_{I}-2\Omega_{T}\right)
\end{split}
,
\label{res19}
\end{equation}
where $\Phi$ are functions of the semi-major axis ratio only involving the Laplace coefficients and their derivatives.
\begin{table*}
\centering
\begin{tabular}{c|c}
\hline\hline
$\Phi$&Expression\\
\hline
$\Phi_{19, 0}$&$-\left(\frac{99}{8}\alpha^{2}+\frac{63}{8}\alpha^{3}D+\frac{9}{16}\alpha^{4}D^{2}\right)b_{5/2}^{2}$\\$\Phi_{19, 1}$&$\left(\frac{99}{8}\alpha^{2}+\frac{63}{8}\alpha^{3}D+\frac{9}{16}\alpha^{4}D^{2}\right)b_{5/2}^{2}+\left(\frac{135}{8}\alpha^{3}+\frac{15}{2}\alpha^{4}D+\frac{15}{32}\alpha^{5}D^{2}\right)\left(3b_{7/2}^{1}+7b_{7/2}^{3}\right)$\\$\Phi_{19, 2}$&$-\left[\left(\frac{405}{16}\alpha^{3}+\frac{45}{4}\alpha^{4}D+\frac{45}{64}\alpha^{5}D^{2}\right)\left(2b_{7/2}^{1}+5b_{7/2}^{3}\right)+ \left(\frac{1365}{32}\alpha^{4}+\frac{945}{64}\alpha^{5}D+\frac{105}{128}\alpha^{6}D^{2}\right)\left(3b_{9/2}^{0}+19b_{9/2}^{2}+13b_{9/2}^{4}\right)\right]$\\$\Phi_{19, 3}$&$\left(\frac{1365}{32}\alpha^{4}+\frac{945}{64}\alpha^{5}D+\frac{105}{128}\alpha^{6}D^{2}\right)\left(3b_{9/2}^{0}+20b_{9/2}^{2}+14b_{9/2}^{4}\right)$\\
\hline
\end{tabular}
\caption{Semi-major axis functions for the term in equation \ref{res19}.}
\end{table*}
%\paragraph{Initial conditions}
%For every simulation, initial conditions was chosen so that Titan and Iapetus would evolve through a 5:1 resonance during the simulation. In order to accomplish this task, Titan's semi-major was taken to be smaller than the actual one and set to $\num{8.131e-3}$ astronomical units (initial period ratio $\approx 5.01$) . The satellite would then increase this value through tidal migration until the ratio of the satellites orbital periods approaches 5. On the other side Iapetus' semi-major was chosen to be the one of today. Iapetus starts with a circular orbit and its orbital plane corresponds to the Laplace plane.
\section{Simulation results}
\begin{figure}
\begin{center}
\includegraphics[width=0.5\linewidth]{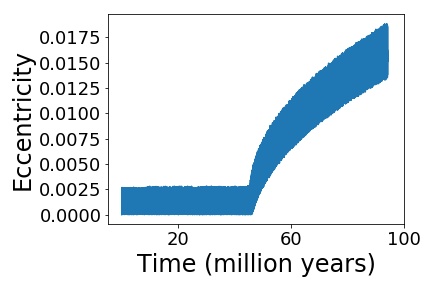}\includegraphics[width=0.5\linewidth]{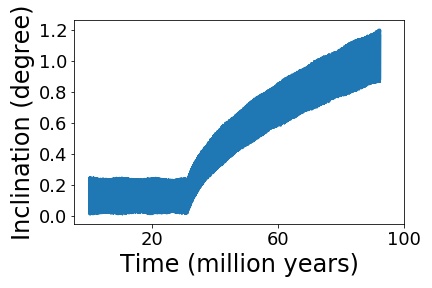}
\caption{Smooth evolutions of the eccentricity and the inclination coming from two different simulations. On the left, the eccentricity of Iapetus growths smoothly after both satellites enter the resonance, while the tilt stays constant. The right figure depicts the evolution of the inclination with respect to the Laplace plane. For this simulation, the eccentricity does not change.}
\label{fig:smooth}
\end{center}
\end{figure}

\begin{figure}
\centering
\includegraphics[width=0.5\linewidth]{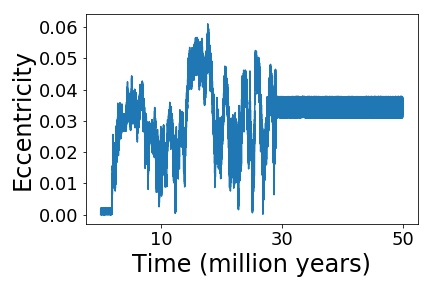}\includegraphics[width=0.5\linewidth]{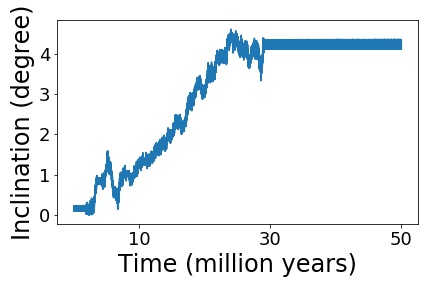}
\caption{Example figures of outputs obtained from our simulations. In this example we clearly distinguish a regular evolution before the resonance, a chaotic resonance crossing, and finally regular dynamics after the resonance. In this specific example, the eccentricity increase is around $0.035$ and the tilt has grown to around $4.5$. These two figures were taken from a run of 100 simulations using the semi-analytic model with $Q=200$.}
\label{fig:decap}
\end{figure}

In this section, we show several results coming from numerical simulations. The eccentricity and the inclination can both be affected by the mean motion resonance crossing in general. Both elements can behave chaotically during the crossing and their resulting values are unpredictable. On the other hand, simulations have shown "smooth" resonance captures, in which either the eccentricity or the inclination evolve smoothly (Figure \ref{fig:smooth}). The great majority of them came from the N-body code and an analysis of all simulations shows that this behaviour happens more often for high values of $Q$ (Table \ref{smootht}). However, because only one element evolves we have discarded these evolutions as a possible scenario.
\begin{table}
\centering
\begin{tabular}{|c|c|}
\hline
$Q$ & Number of smooth simulations\\
\hline
\hline
20&0\\
\hline
100&1\\
\hline
200&13\\
\hline
400&23\\
\hline
600&26\\
\hline
1500&41\\
\hline
\end{tabular}
\caption{Numbers of smooth evolutions as a function of $Q$ using the N-Body code. For each value of $Q$, 100 simulations have been produced}
\label{smootht}
\end{table}
We have concentrated our analysis on the chaotic evolutions and one can distinguish two different types:
\begin{itemize}
\setlength{\itemsep}{1\baselineskip}
\item Fast crossings, for $Q$ under 100. Usually, Titan rushes through the resonance and this prevents any capture. Still, the effect on the eccentricity of Iapetus can result in a kick of a few percent, reaching sometimes $0.05$. On the other hand, the inclination is not really affected at this point. A few degrees can be obtained in the best cases.
\item For values of $Q$ over 100, the majority of simulations show a capture and a chaotic evolution of Iapetus' orbit. Then, several scenarios can take place. Either the capture is maintained and its eccentricity grows until Iapetus is ejected from the system; or Iapetus is released from the resonance and continues on a regular orbit with an excited eccentricity and inclination as shown in Fig. \ref{fig:decap}.
\end{itemize}
%\begin{figure}[H]
%\begin{center}
%\begin{subfigure}{0.7\textwidth}
%\centering
%\includegraphics[width=1\linewidth]{Q20_cart_laplace.jpg}
%\end{subfigure}
%\begin{subfigure}{1\textwidth}
%\centering
%\includegraphics[width=0.5\linewidth]{M830eccentricity.jpg}\includegraphics[width=0.5\linewidth]{M830inclination.jpg}
%\end{subfigure}
%\caption{Bottom figures come from an example of a resonance crossing with $Q=20$ using the N-Body code. The obvious effect of the resonance is here a "kick" in both eccentricity and free inclination. The histogram at the top indicates the orbital growth of Iapetus' element during the crossing.}
%\end{center}
%\end{figure}

\begin{figure}
\centering
\includegraphics[width=0.5\linewidth]{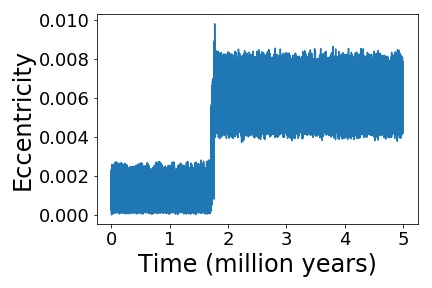}\includegraphics[width=0.5\linewidth]{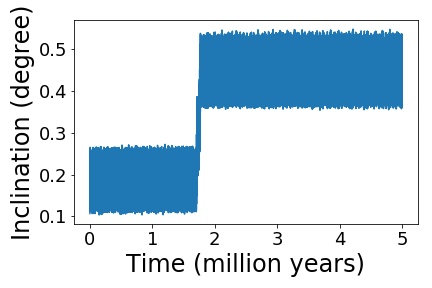}
\caption{Example of a resonance crossing with $Q=50$ using the N-Body code. The obvious effect of the resonance is here a "kick" in both eccentricity and free inclination.}
\end{figure}
\begin{figure}
\resizebox{\hsize}{!}{\includegraphics{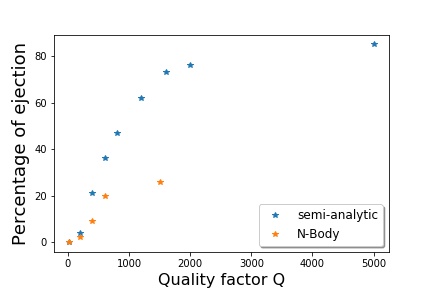}}
\caption{Number of ejections as a function of the effective quality factor of the planet. Each dot is representative of a run of 100 simulations done with the same value of Q. We observe that the number of ejections using the semi-analytic model is greater than using the N-body. The first reason is that the percentages are computed with all simulations considered, smooth evolutions included. The difference then gets more and more pronounced as the number of smooth evolutions also increases with $Q$ for the N-Body code (Table \ref{smootht}), but stays rather low for the semi-analytic simulations. The reason for such differences is probably that the rates of mean longitudes are not properly modelled when we average out all the terms except the resonant ones and such bias appears during the resonance where mean longitudes play an important role. Also for chaotic simulations, the averaged equations are truncated to the fourth power, meaning that they are not suited for high eccentricities. Iapetus nearly never survives if its eccentricity grows over $0.25$ during the resonance. Therefore the threshold for the semi-analytic code was set to $0.25$ for Iapetus to be considered ejected.}
%The right figure illustrates the chaotic behaviour of Iapetus' eccentricity before getting ejected and like any ejections observed with the N-Body code,
\label{fig:ejectionQ}
\end{figure}

\begin{figure}
\centering
\includegraphics[width=0.5\linewidth]{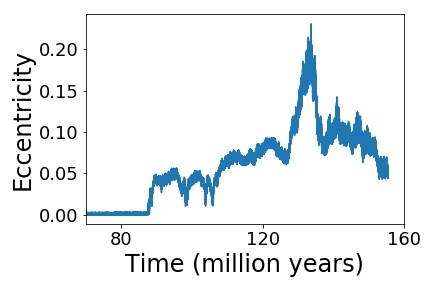}\includegraphics[width=0.5\linewidth]{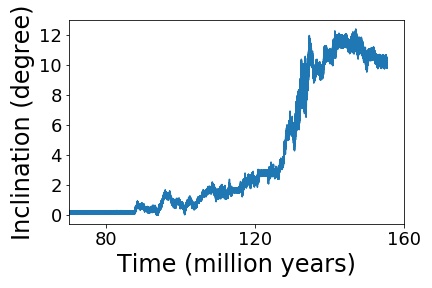}
\caption{Example of a simulation showing post-resonance eccentricity and inclination compatible with Iapetus' actual orbit. Here, the resonance is crossed 90 millions years after the start of the simulation. As Iapetus evolves chaotically during the crossing, its eccentricity rises over $0.2$, while the tilt also increases to over 10 degrees. On such an orbit, Iapetus is expected to get ejected, but at the end, before getting out of the resonance, the eccentricity decreases to $0.05$ while the tilt stays well over the value we observe today.}
\label{fig:hugeei}
\end{figure}

Usually, Iapetus never comes out of the resonance with its eccentricity being over $0.25$. However, it can still happen that the eccentricity increases over this upper limit while both satellites are still in resonance. When it happens, the fate of Iapetus will simply be an ejection of the system (eccentricity rises over 1). A quick analysis of all simulations made with different values of $Q$ is sufficient to identify the following behaviour: the number of ejections increases with $Q$. In other words, the slower Titan passes through the resonance, the more Iapetus is likely to be ejected, as shown in Fig. \ref{fig:ejectionQ}. This is true for both models. For the N-Body code, one can extrapolate that ejections and smooth evolutions will be dominant for very high values of $Q$ (here between $2\,000$ and $18\,000$), while for $Q$ between 100 and 2000, de-capture happens in a dominant fashion. Being less precise than the direct method, the semi-analytic code produces less smooth captures but also shows similar outcomes concerning ejections. We observe that the number of ejections is greater for the semi-analytic  than for the N-Body code (Figure \ref{fig:ejectionQ}) but besides this difference, post resonance eccentricities and inclinations are quite similar.
%Usually, Iapetus never comes out of the resonance with its eccentricity being over $0.20$. However, it can still happen that the eccentricity increases over this upper limit while both satellites are still in resonance. When it happens, the fate of Iapetus will simply be an ejection of the system (eccentricity rises over 1), as shown in \autoref{fig:ejectionQ}. A quick analysis of all simulations made with different values of $Q$ is sufficient to identify the following behaviour : the number of ejections increases with $Q$. In order words, the slower Titan passes through the resonance, the more Iapetus is likely to get ejected. Simulations show that at very high values of Q (here between 2000 and 18000), the ejection is almost certain, while we can find a wide range of outcome for Q between 100 and 2000.
However, it is in the range between 100 and 2000 that simulations show the best agreement between final elements and the orbit of Iapetus as we see it today. For chaotic evolutions, if Iapetus is not ejected, it will come out of the resonance with a tilt to its local Laplace plane of a few degrees, usually spanning from 0 to 5 degrees, but several simulations have shown free inclination up to 11 degrees. We have to point out that those huge inclination excitations are quite rare (less than 1\% of simulations) and also generally come with high eccentricities. Out of 1400 simulations done with both codes with $Q$ between 100 and 2000, we note that 11 simulations show a free inclination over 5 degrees, nine of them have an eccentricity over 5\%, and two come with a lower eccentricity (Fig. \ref{fig:hugeei} and another showing a tilt of around 7 degrees and 1\% eccentricity). On top of this, 57 of them have a free inclination over 3 degrees with an eccentricity spanning from 0 to 15\%. \footnote{For example, one simulation reaches 10.6 degrees with almost 0.1 eccentricity while another shows 4.8 degrees and 0.038 in eccentricity.} Both elements have a correlated evolution, even though they evolve chaotically. High free inclinations usually come out with the eccentricity in excess (over 0.05), but the very low number of trajectories compatible with Iapetus' orbital elements of today still gives us confidence that Titan could have still excited Iapetus' orbit to its current behaviour.
\section{Discussion on the model}
We have left aside the gravitational effect of the outer planets, but it is worth asking if it is a fair negligence. Jupiter evolves close to 5:2 mean motion commensurability with Saturn and plays a major role in its secular motion around the Sun. One of the main aspects of this perturbation is a secular change of Saturn's eccentricity with a period of around 50 000 years (\cite{murray2000solar}). If we go back to Eq. \ref{fee}, we see that the Laplace plane equilibrium has a dependency on the planet's eccentricity (alias the Sun's eccentricity in our model). However, although the Laplace plane equilibrium will change at the same frequency as the Sun's eccentricity, the free inclination will still preserve a constant tilt to it.

The precession of the spin axis of Saturn was also a subject of concern to obtain the correct dynamics. We have taken it into consideration and implemented a code where both satellites orbit a precessing Saturn. Numerical simulations show that the satellite orbital planes will follow the precessing spin pole of the planet. This dynamic was also studied in the pioneer paper \cite{1965AJ.....70....5G} in which the perturbing effects were those of the precession of the spin axis of the planet and its flattening. The slow precession of Saturn (\cite{2017Icar..290...14F}) acts like an adiabatic process that does not alter the mutual perturbation between satellites.
%Here, although we have added the mutual gravitational effects of both satellites and also the Sun, both satellite orbits keep a similar evolution.
%Using different averaged models and augmented numerical precision

Hyperion was also integrated with the other bodies to monitor any significant changes to its orbit and to the 4:3 mean motion resonance with Titan. No radical changes were detected during the 5:1 crossing between Titan and Iapetus. Hyperion would have therefore easily survived a triple commensurability with Titan and Iapetus; this result was already outlined in a previous study (\cite{2013arXiv1311.6780C}). We have also found that the commensurability is preserved regardless of the tidal migration.
%Hyperion's orbit can be used to constrain the past evolution of Titan. Studies have shown that Titan's eccentricity can not have been over 6-$7\%$ for the capture to take place. Also the eccentricity of Hyperion (today around $10\%$) [Colombo] [Sinclair]

In this study we have neglected tides acting on satellites but the question of damping the eccentricity of Iapetus arises here because of the great number of simulations showing high post-resonance eccentricities. For tides acting on a satellite, we have \citep{1991Icar...94..399M}
\begin{equation}
\frac{de}{dt}=-\frac{21}{2}\frac{k_{2,s}nm_{S}}{Q_{s}m_{s}}\left(\frac{R_{s}}{a}\right)^{5}e
,\end{equation}
where $k_{2,s}$ and $Q_{s}$ are the Love number and the quality factor of the satellite, $m_{s}$ and $R_{s}$ its mass and radius and $m_{S}$ the mass of Saturn. Such an equation admits an evolution timescale, in years, of
\begin{equation}
\tau\approx 25\times10^{9}\frac{Q_{s}}{k_{2,s}}
,\end{equation}
which is well over the age of the solar system. Therefore, tides are too weak to damp the satellite's eccentricity during its excitation. Still it is worth mentioning that stronger tidal damping of Iapetus' eccentricity should occur through its secular coupling to Titan's eccentricity \citep{2014AJ....148...52N}.

The semi-analytic model was also tested using quadrupole precision floating-point numbers and has shown similar outcomes to simulations made in double precision. Also more solar terms involving the Sun's mean longitude were added as well as terms associated to the relation $10\lambda_{Iapetus}-2\lambda_{Titan}$, but no real new behaviour appeared.
\section{Conclusion}
We have simulated a mean motion resonance crossing between Titan and Iapetus using direct N-body integration as well as averaged equations of motion. Both methods have shown similar behaviour regarding statistical outcomes for the fate of Iapetus. The averaged equations were used to study the crossing for slow migration rates of Titan and, as for the N-Body code, showed a large number of ejections of Iapetus if $Q$ was set to a value over 2000. A fast migration rate would preserve Iapetus' eccentricity under a few percent, without however disturbing its orbital tilt to the Laplace plane. The rise of eccentricity is generally consistent with today's value of around $3\%$ for $Q$ between 20 and 2000. Those values are consistent with quality factors found for the icy satellites in \cite{2012ApJ...752...14L} and \cite{2017Icar..281..286L}. Simulations with $Q$ set over 2000 have shown a majority of unwanted scenarios such as ejections or smooth evolutions, therefore $Q=2000$ sets a lower limit to the tidal energy dissipation acting inside Saturn. A quality factor over 100 is necessary to raise Iapetus' free inclination to values approaching the tilt to the Laplace plane observed today. The tilt mainly stays beneath 5 degrees, but a few simulations have shown tilts to the Laplace plane reaching 8 degrees and beyond with an eccentricity relatively small. Therefore, although the eccentricity excitation of Iapetus is easy to obtain with this resonance, increasing the free inclination is harder. However, a few simulations (less than 1\%) have shown some compatibility with Iapetus' orbit, therefore the scenario that Titan is responsible for Iapetus' orbit with this past resonance is plausible (cf \ref{fig:hugeei}). A quality factor at Titan's frequency between 100 and 2000 implies that the resonance crossing happened between 40 million and 800 billion years ago. This result reinforces the idea of strong energy dissipation inside Saturn.
%An alternative and more likely scenario would explain Iapetus' eccentricity with this resonance but would assume a primordial tilt to the Laplace Plane. In that case, the value of $Q$ could be around 20, which is in the range of values given by the theory in \cite{2016MNRAS.458.3867F}.
\begin{acknowledgements}
The authors acknowledge the support of the contract Prodex CR90253 from the Belgian Science Policy Office (BELSPO).
This study is part of the activities of the ISSI Team \emph{The ENCELADE Team: Constraining the Dynamical Timescale and Internal Processes of the Saturn and Jupiter Systems from Astrometry}.
Computational resources have been provided by the Consortium des \'Equipements de Calcul Intensif (C\'ECI), funded by the Fonds de la Recherche Scientifique de Belgique (F.R.S.-FNRS) under Grant No. 2.5020.11. Also, this work was granted access to the HPC resources of MesoPSL financed by the Region Ile de France and the project Equip@Meso (reference ANR-10-EQPX-29-01) of the programme Investissements d'Avenir supervised by the Agence Nationale pour la Recherche. VL's research was supported by an appointment to the NASA Postdoctoral Program at the NASA Jet Propulsion Laboratory, Caltech, administered by the Universities Space Research Association
under contract with NASA. The authors are indebted to all participants of the ISSI Encelade team.
\end{acknowledgements}

%----------------------------------------------------------------------------------------
%       REFERENCE LIST
%----------------------------------------------------------------------------------------

\bibliographystyle{aa}
\bibliography{bibliography}

%----------------------------------------------------------------------------------------

\end{document}